\documentstyle[epsfig,aps]{revtex}

\newcommand{\ccap}[2]{\caption[#1]{#2}\label{#1}\vspace{0.2cm}}
\begin{document}
\draft
\title{ Anomalous WW$\gamma$ Vertex in  
$\gamma p$ Collision}
\mediumtext
\author{S. Ata\u{g}\footnotemark and \.{I}.T. \c{C}ak\i r} 
\address{Ankara University Department of Physics \\
 Faculty of Sciences, 06100 Tandogan, Ankara, Turkey}
\footnotetext{Corresponding author: Fax:+90 312 2232395, 
e-mail: atag@science.ankara.edu.tr}
\maketitle

\begin{abstract}
The potential of LC+HERAp based $\gamma$p collider to probe 
WW$\gamma$ vertex is presented through the discussion of 
sensitivity to anomalous couplings and $p_{T}$ distribution
of the final quark. The limits of  $-0.04<\Delta\kappa<0.04, 
\;\;\; -0.11<\lambda<0.11, \;\;\ $ at 95\% C.L. can be reached 
with integrated luminosity 200$pb^{-1}$. The limit for 
$\Delta\kappa$ is comparable to one which is expected from LHC.   
The bounds  are also obtained from  corresponding ep collider using 
Weizs\"{a}cker-Williams Approximation to compare with real photons.  
\end{abstract}

\vskip 0.5cm

\section{Introduction}
Recently there have been intensive studies to test the 
deviations from the Standard Model (SM) at present and 
future colliders. The investigation of three gauge boson 
couplings plays an important role to manifest the non 
abelian gauge symmetry in standard electroweak theory.
The precision measurement of the triple vector boson 
vertices will be the crucial test of the structure of the 
SM. 

It is known that present collider measurements 
at LEP2\cite{lep} and Tevatron\cite{tevatron} 
can not  probe anomalous triple gauge boson self couplings to 
precision better than O($10^{-1}$). 
Further analyses of $WW\gamma$ vertex has been given by several 
papers  for ep collider HERA
\cite{dubinin,baur1,kim,baur2,janssen,noyes}, 
projected future colliders LHC\cite{atlas}  and 
linear electron-positron collider (LC)\cite{desy}.
After  LC is constructed $\gamma e$, $\gamma\gamma$,
linac-ring type $ep$ and $\gamma p$ modes should be discussed
and may work as complementary to basic colliders. 
The Linear Collider 
 design at DESY \cite{desy,trines} is the only one
 that can be converted into an ep \cite{brinkmann,roeck} collider.
 $\gamma p$ collider mode is an additional advantage of 
this linac-ring type ep collider \cite{ciftci,zaydin} where 
real gamma beam is obtained by the 
Compton backscattering of laser
 photons off linear electron beam. Estimations show
 that the luminosity for $\gamma$p collision turns out to be of
 the same order as the one for ep collision 
due to  the fact that
$\sigma_{p}>>\sigma_{\gamma}$ where $\sigma_{p}$ and 
$\sigma_{\gamma}$ are transverse sizes of proton 
and photon bunches at collision point. Since
most of the photons are produced at high energy region in the Compton
backscattering  the cross sections are about one order of magnitude 
larger than parental ep collider for photoproduction processes.
 According to present project at
DESY 500 GeV electrons from linear electron beam are allowed to collide
820 GeV protons from HERA ring\cite{brinkmann,roeck}.  
Parameters of this LC+HERAp  and its  
$\gamma$p option are shown in Table \ref{tab1} \cite{ciftci}. 
Therefore such kind of high energy $\gamma$p colliders will possibly  
give additional information  to linac-ring type ep colliders for a
variety of processes\cite{zaydin,auhep}.
In this paper we examine the potential of
future LC+HERAp based $\gamma$p collider to probe anomalous WW$\gamma$
coupling and compare the results with those from its basic ep collider 
and other projected future colliders.

\begin{table}[bth]
\caption{Main parameters of LC+HERAp based ep and  $\gamma$p
colliders.
\label{tab1}}
\begin{center}
\begin{tabular}{lccc}
Machine &$\sqrt{s_{ep}}$ TeV & $\sqrt{s_{\gamma p}}^{max}$ TeV
&$L_{ep}\simeq L_{\gamma p}(cm^{-1}s^{-1})$   \\
\hline
LC+HERAp  &1.28 &1.16 & 1$\times 10^{31}$ \\
\end{tabular}
\end{center}
\end{table}

\section{ Lagrangian and Cross Sections}

C and P parity conserving effective lagrangian for two charged
W-boson and one photon interaction can be written following the papers
\cite{gaemers,hagiwara}

\begin{eqnarray}
L&&=e(W_{\mu\nu}^{\dagger}W^{\mu}A^{\nu}-
W^{\mu\nu}W_{\mu}^{\dagger}A_{\nu}+
\kappa W_{\mu}^{\dagger}W_{\nu}A^{\mu\nu} 
+{\lambda\over M_{W}^{2}}W_{\rho\mu}^{\dagger}
W_{\nu}^{\mu}A^{\mu\rho})
\end{eqnarray}
where

\begin{eqnarray}
W_{\mu\nu}=\partial_{\mu}W_{\nu}-\partial_{\nu}W_{\mu}\nonumber 
\end{eqnarray}
and dimensionless parameters $\kappa$ and $\lambda$ are related
to the magnetic dipol and electric quadrupole moments.
For $\kappa=1$ and $\lambda=0$  Standard Model is restored.
In momentum space this has the following form with momenta
$W^{+}(p_1)$,$W^{-}(p_2)$ and $A(p_3)$

\begin{eqnarray}
\Gamma_{\mu\nu\rho}(p_{1},p_{2},p_{3})=&&e[g_{\mu\nu}
(p_{1}-p_{2}-{\lambda\over M_{W}^{2}}[(p_{2}.p_{3})p_{1}
-(p_{1}.p_{3})p_{2}])_{\rho}\nonumber \\
&&+g_{\mu\rho}(\kappa p_{3}-p_{1}
+ {\lambda\over M_{W}^{2}}[(p_{2}.p_{3})p_{1}
-(p_{1}.p_{2})p_{3}])_{\nu} \nonumber \\
&&+g_{\nu\rho}(p_{2}-\kappa p_{3}
-{\lambda\over M_{W}^{2}}[(p_{1}.p_{3})p_{2}
-(p_{1}.p_{2})p_{3}])_{\mu} \nonumber \\
&&+{\lambda\over M_{W}^{2}}(p_{2\mu}p_{3\nu}p_{1\rho}
-p_{3\mu}p_{1\nu}p_{2\rho}])]
\end{eqnarray}
where  $p_1+p_2+p_3=0$.
There are three Feynman diagrams for the subprocess
$\gamma q_{i}\rightarrow Wq_{j} $ and  only t-channel W
exchange graph contributes $WW\gamma$ vertex. One should
note that $\gamma p$ collision isolates $WW\gamma$ coupling but
many  processes in $e^{+}e^{-}$, pp and ep collisions include 
mixtures of $WW\gamma$ and WWZ couplings.

The unpolarized differential cross section for the subprocess
$\gamma q_{i}\rightarrow Wq_{j} $ can be obtained using
helicity amplitudes from \cite{baur2} summing over the helicities

\begin{eqnarray}
{d\hat{\sigma}\over d\hat{t}}={2\over{\hat{s}-M_{W}^{2}}}
{\beta\over {64\pi\hat{s}}} \sum_{\lambda_{\gamma}
\lambda_{W}}{1\over 2} M_{\lambda_{\gamma}\lambda_{W}}^{2}
\end{eqnarray}
where
\begin{eqnarray}
 M_{\lambda_{\gamma}\lambda_{W}}={e^{2}\over{\sqrt{2}
 \sin\theta_{W}}}{\hat{s}\over{\hat{s}+M_{W}^{2}}}
\sqrt{\beta}A_{\lambda_{\gamma}\lambda_{W}}\; , \;\;\;\; 
\beta=1-{M_{W}^{2}\over\hat{s}} 
\end{eqnarray}
and $\theta_{W}$ is the Weinberg angle. 

For the signal we are considering a quark jet and on-shell
W with leptonic decay mode
\begin{eqnarray}
\gamma p\rightarrow W^{\mp}+jet\rightarrow\ell+p_{T}^{miss}+jet \; ,
\;\;\;\;\;\; \ell=e,\mu 
\end{eqnarray}
In this mode charged lepton and the quark jet are in general well
separated and the signal is in principle free of background of SM.

The total cross section for the subprocess 
$ \gamma q_{i} \rightarrow W q_{j}$ can be divided into two parts,
direct, and resolved-photon production 
\begin{eqnarray}
\hat{\sigma}=\hat{\sigma}_{dir}+\hat{\sigma}_{res}
\end{eqnarray}
The direct part is given as follows 

\begin{eqnarray}
\hat{\sigma}=&&{{\alpha G_{F}M_{W}^{2}}\over{\sqrt{2}\hat{s}}}
|V_{q_{i}q_{j}}|^{2}\{(|e_{q}|-1)^{2}(1-2\hat{z}+2\hat{z}^{2})
\log({\hat{s}-M_{W}^{2}\over\Lambda^{2}})
-[(1-2\hat{z}+2\hat{z}^{2}) \nonumber \\
&&-2|e_{q}|(1+\kappa+2\hat{z}^{2})
+{{(1-\kappa)^{2}}\over{4\hat{z}}}-{{(1+\kappa)^{2}}\over{4}}]
\log{\hat{z}}+ [(2\kappa+{{(1-\kappa)^{2}}\over{16}})
{1\over \hat{z}} \nonumber \\
&&+({1\over 2}+{{3(1+|e_{q}|^{2})}\over{2}})\hat{z}
+(1+\kappa)|e_{q}|-{{(1-\kappa)^{2}}\over{16}}
+{|e_{q}|^{2}\over 2}](1-\hat{z}) \nonumber \\
&&-{{\lambda^{2}}\over{4\hat{z}^{2}}}(\hat{z}^{2}-2\hat{z}
\log{\hat{z}}-1)+{{\lambda}\over{16\hat{z}}}
(2\kappa+\lambda-2)[(\hat{z}-1)(\hat{z}-9)
+4(\hat{z}+1)\log{\hat{z}}]\}
\end{eqnarray}
where $\hat{z}=M_{W}^{2}/\hat{s}$ and $\Lambda^{2}$ is
cut off scale in order to regularize $\hat{u}$-pole of the
colinear singularity for massles quarks. In the case of massive
quarks there is no need such a kind of  cut. 
The resolved-photon production cross section can be calculated 
using Breit-Wigner formula for $ q_{\gamma}q_{p}\to W$ fusion process
\begin{eqnarray}
\hat{\sigma}(q_{i}\bar{q}_{j}\to W)={{\pi\sqrt{2}}\over{3}}G_{F}m_{W}^{2}
|V_{ij}|^{2}\delta(x_{i}x_{j}s_{\gamma p}-m_{W}^{2})
\end{eqnarray}
where $V_{ij}$ is the Cabibbo-Kobayashi-Maskawa (CKM) matrix.
For the integrated cross sections we need parton distribution
functions inside the photon and proton. The photon structure 
function $f_{q/\gamma}$ consists of perturbative pointlike parts 
and hadronlike parts. The pointlike part can be calculated in the 
leading logarithmic approximation and is given by the expression

\begin{eqnarray}
f_{q/\gamma}^{LO}(x,Q_{\gamma}^{2})={{3\alpha e_{q}^{2}}\over{2\pi}}
[x^{2}+(1-x)^{2}]\log{Q_{\gamma}^{2}\over\Lambda^{2}} 
\end{eqnarray} 
where $e_{q}$ is the quark charge. 
For the integrated total cross section over the quark distributions 
inside the proton, photon and the spectrum of backscattered laser 
photon the following result can be obtained easily 

\begin{eqnarray}
\sigma_{res}=\sigma_{0}\int_{m_{W}^{2}/s}^{0.83}dx 
\int_{x}^{0.83}{{dy}\over{xy}}f_{\gamma/e}(y)f_{q_{i}/p}
({{m_{W}^{2}}\over{xs}},Q_{p})[f_{q_{j}/\gamma}({{x}\over{y}},
Q_{\gamma}^{2})-f_{q_{j}/\gamma}^{LO}({{x}\over{y}},Q_{\gamma}^{2})] 
\end{eqnarray}
with
\begin{eqnarray}
\sigma_{0}={{\pi\sqrt{2}}\over{3s}}G_{F}m_{W}^{2}
|V_{ij}|^{2}
\end{eqnarray}

Since the contribution from the pointlike part of the 
photon structure function  was already taken into 
account in the calculation of the direct part 
it was subtracted from 
$f_{q/\gamma}(x,Q_{\gamma}^{2})$ in the above 
formula to avoid double counting on the leading 
logarithmic level. 

In a similar way direct part of the  integrated cross section 
can be obtained 

\begin{eqnarray}
\sigma_{dir}=\int_{\tau_{min}}^{0.83}d\tau \int_{\tau/0.83}^{1}
{dx\over x}f_{\gamma/e}(\tau/x)f_{q/p}(x,Q^{2})\hat{\sigma}(\tau s)
\end{eqnarray}
with $\tau_{min}=(M_{W}+M_{q})^{2}/s$

The sum of resolved and direct contribution, in principle, should 
not depend on value of the parameter $\Lambda$.
The effects of $\Lambda$ on the cross sections and 
contributions of direct and resolved parts  are given in 
Table \ref{tab2}. 
In Table \ref{tab3} integrated total cross sections times branching
ratio of $W\rightarrow \mu\nu $ and corresponding number of events 
 are shown for various values of
$\kappa$  and $\lambda$. Through the calculations proton 
structure functions of Martin, Robert and
Stirling (MRS A)\cite{martin} and  photon structure functions 
of Drees and Grassie (DG)\cite{drees} have been used
with $Q^{2}=M_{W}^{2}$.
Here number of events has been calculated using 
\begin{eqnarray}
N=\sigma(\gamma p\rightarrow W+Jet)B(W\rightarrow \mu\nu)
A L_{int}
\end{eqnarray}
where  we take the acceptance in the muon channel A  and integrated 
luminosity $L_{int}$ as  65\% and 200pb$^{-1}$.
To give an idea about the comparison with corresponding ep
collider the cross sections obtained using Weizs\"{a}cker-Williams
approximation are also shown on the same table. 
As the cross section $\sigma(\gamma p\rightarrow W+Jet)$ we have
considered the sum of $\sigma(\gamma u\rightarrow W^{+}d)$,
$\sigma(\gamma \bar{d}\rightarrow W^{+}\bar{u})$,
$\sigma(\gamma \bar{s}\rightarrow W^{+}\bar{c})$,
$\sigma(\gamma u\rightarrow W^{+}s)$,
$\sigma(\gamma \bar{s}\rightarrow W^{+}\bar{u})$, and
$\sigma(\gamma \bar{d}\rightarrow W^{+}\bar{c})$.
As shown from Table \ref{tab3} the cross sections using backscattered
laser photons are considerably larger than the case of corresponding
ep collision. We assume that the form factor structure of
$\kappa-1$ and $\lambda$ do not depend on the momentum transfers
at the energy region considered.
Then anomalous terms containing
$\kappa$ grow with $\sqrt{\hat{s}}/M_{W}$ and terms with $\lambda$
rise with $\hat{s}/M_{W}^{2}$. 
Deviation $\Delta\kappa=\kappa-1=1$  from SM value changes the total
cross sections 30-70\% whereas the $\Delta\lambda=\lambda=1$ gives
80\% changes. Therefore high energy will improve the sensitivity
to anomalous couplings. For comparison with HERA energy
$\sqrt{s}=314$ GeV the similar results would be 20-40\% for
$\Delta\kappa=1$ and 5\% for $\Delta\lambda=1$.

\begin{table}[bth]
\caption{ Integrated total cross section times branching ratios
$\sigma(\gamma p\rightarrow Wj)\times B(W^{+}\rightarrow
 \mu\nu)$ in pb to indicate direct and resolved photon
contribution depending on invariant kinematical cut $\Lambda$ in GeV.
DG and MRS A were used for photon and proton structure functions.
\label{tab2}}
\begin{center}
\begin{tabular}{clll}
    & $\sigma_{dir}\times B(W^{+}\rightarrow \mu\nu)$& &  \\
\hline
$\kappa$, $\lambda$ & $\Lambda=0.2$ & $\Lambda=1$ & $\Lambda=5$\\
\hline
1,0 & 12.12 & 11.34 & 10.57 \\
1,1  &23.43 &22.66 &21.88 \\
1,2 &57.37 &56.59 &55.82  \\
0,0 &8.08 &7.31 &6.53  \\
2,0 &21.70  &20.92  &20.14 \\
\hline
 &  $\sigma_{res}\times B(W^{+}\rightarrow \mu\nu)$&  & \\
\hline
 & -0.77  & 0.92  & 2.62   \\
\end{tabular}
\end{center}
\end{table}

\begin{table}[bth]
\caption{Integrated total cross section times branching
ratio $\sigma(\gamma p\rightarrow Wj)\times B(W^{+}\rightarrow
\mu\nu)$ in pb and corresponding number of events(in parentheses)
without cutoff for massive quarks. Integrated luminosity
$L_{int}=200 pb^{-1}$ has been used.
\label{tab3}}
\begin{center}
\begin{tabular}{llcc}
Backscattered Laser  &  WWA & $\kappa$  & $\lambda $ \\
\hline
13.8(1780) &1.3(170) &1 & 0 \\
25.1(3262) &1.9(246) &1 & 1 \\
59.0(7670) &3.7(480) &1 & 2 \\
9.7(1262) & 0.9(120) &0 & 0 \\
23.4(3042) &2.1(274) &2 & 0
\end{tabular}
\end{center}
\end{table}

It is important to see how the anomalous couplings change the 
shape of the transverse momentum distributions of the final quark 
jet. For this reason we  use  the following formula:

\begin{eqnarray}
{d\sigma\over dp_{T}}=&&2p_{T}\int_{y^{-}}^{y^{+}}dy
\int_{x_{a}^{min}}^{0.83}dx_{a} f_{\gamma/e}(x_{a})
f_{q/p}(x_{b},Q^{2}) 
({x_{a}x_{b}s \over {x_{a}s-2m_{T}E_{p}e^{y}}})
{d\hat{\sigma}\over d\hat{t}} 
\end{eqnarray}
where
\begin{eqnarray}
y^{\mp}=\log{[{{0.83s+m_{q}^{2}-M_{W}^{2}}
\over {4m_{T}E_{p}}}\mp\{ ({{0.83s+m_{q}^{2}-M_{W}^{2}}
\over {4m_{T}E_{p}}})^{2}-{0.83E_{e}\over E_{p}}\}^{1/2}]}
\end{eqnarray}
\begin{eqnarray}
&&x_{a}^{(1)}={{2m_{T}E_{p}e^{y}-m_{q}^{2}+M_{W}^{2}}
\over{s-2m_{T}E_{e}e^{-y}}} \;\; , \;\;\;\;\;
x_{a}^{(2)}={(M_{W}+m_{q})^{2}\over s} \nonumber \\
&&x_{a}^{min}=MAX(x_{a}^{(1)},x_{a}^{(2)}) \;\; , \;\;\;\;\;
x_{b}={{2m_{T}E_{e}x_{a}e^{-y}-m_{q}^{2}+M_{W}^{2}}
\over {x_{a}s-2m_{T}E_{p}e^{y}}}
\end{eqnarray}
with
\begin{eqnarray}
&&\hat{s}=x_{a}x_{b} \;\; , \;\;\;
\hat{t}=m_{q}^{2}-2E_{e}x_{a}m_{T}e^{-y} \;\; , \;\;\;
\hat{u}=m_{q}^{2}+M_{W}^{2}-\hat{s}-\hat{t} \nonumber \\
&&m_{T}^{2}=m_{q}^{2}+p_{T}^{2}
\end{eqnarray}
The $p_{T}$ spectrum B($W\rightarrow \mu\nu )
\times d\sigma/dp_{T}$ of the quark jet is 
shown in Fig. \ref{fig1} for various $\kappa$ and 
$\lambda$ values at LC+HERAp based $\gamma$p collider.
Similar distribution is given in 
Fig. \ref{fig2} for Weizs\"{a}cker-Williams Approximation that 
covers the major contribution from ep collision.
It is clear that the cross section at 
large $p_{T}$ is quite sensitive to
anomalous $WW\gamma$ couplings. As $\lambda$ increases the cross 
section grows more rapidly when compared with $\kappa$ 
dependence at high $p_{T}$ region $p_{T}>100$ GeV. 
The cross sections with real gamma beam are one
order of magnitude larger than the case of WWA. 
Comparison between two figures also shows that the curves  
become more separable as $\hat{s}$ gets large.     

\begin{figure}[htb]
  \begin{center}
  \epsfig{file=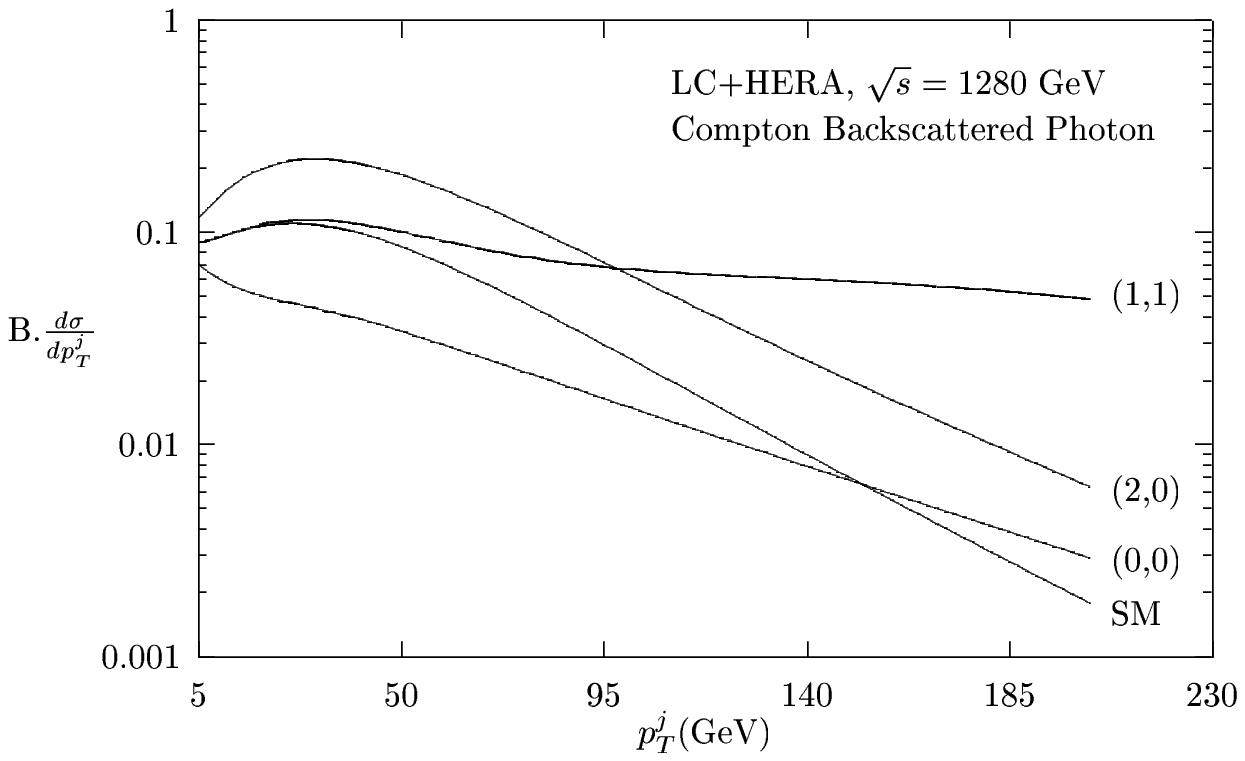}
  \end{center}
  \ccap{fig1}{\footnotesize  $\kappa$ and $\lambda$ dependence
of the transverse momentum distribution of the quark jet at
LC+HERAp based $\gamma$p collider (Compton Backscattered Photon).
The unit of the vertical axis is pb/GeV and the numbers
in the parentheses stand for anomalous coupling parameters
$(\kappa ,\lambda )$.  }
\end{figure}

\begin{figure}[htb]
  \begin{center}
  \epsfig{file=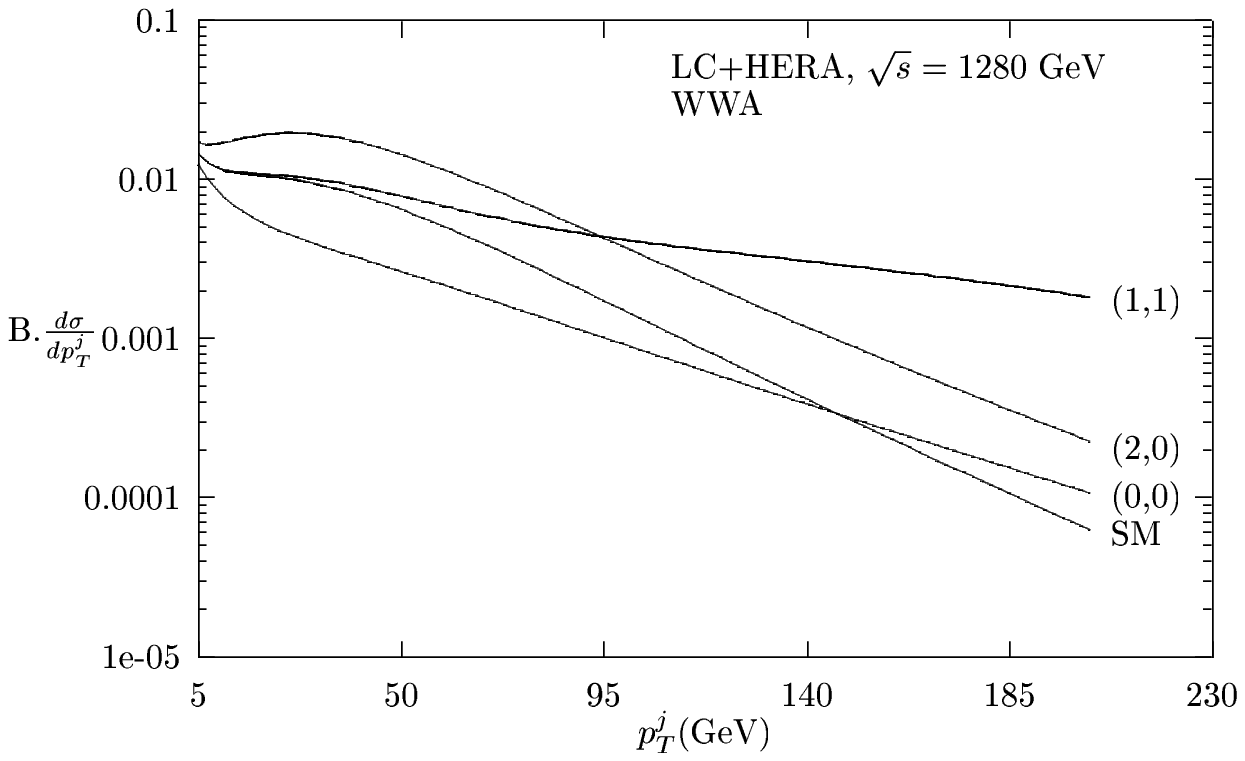}
  \end{center}
  \ccap{fig2}{\footnotesize  The same as the Fig.1 but for
Weizs\"{a}cker Williams Approximation.}
\end{figure}

\section{Sensitivity to Anomalous Couplings}
   
We can estimate sensitivity of LC+HERAp based $\gamma$p collider to
anomalous couplings by assuming the 0.02 combined systematic error
in the luminosity  measurement and detector  acceptance 
for the integrated luminosity value of 200 pb$^{-1}$.
We use the simple $\chi^{2}$-criterion to obtain 
sensitivity as follows

\begin{eqnarray}
\chi^{2}=\sum_{i=bins}({{X_{i}-Y_{i}}\over{\Delta_{i}^{exp}}})^{2}
\end{eqnarray}
\begin{eqnarray}
X_{i}=\int_{V_{i}}^{V_{i+1}} {{d\sigma^{SM}}\over{dV}}dV ,
\;\;\;\;\;\;
Y_{i}=\int_{V_{i}}^{V_{i+1}} {{d\sigma^{NEW}}\over{dV}}dV
\end{eqnarray}

\begin{eqnarray}
\Delta_{i}^{exp}=X_{i}\sqrt{\delta_{stat}^{2}+\delta_{sys}^{2}}
\;,\;\;\;\;\;\;V=p_{T}
\end{eqnarray}
We have divided $p_{T}$ region of the final quark into equal pieces 
for binning procedure and  have considered at least 10 events 
in each bin. For sensitivity, the number of $W^{+}$ and $W^{-}$ 
events given with their branching ratios in the $\mu\nu$ and 
$e\nu$ channels has been taken into account.  
Using the above formula  the  limits on the $\Delta\kappa$
and $\lambda$ are given in Table \ref{tab4}
for the deviation of the  cross section from 
the Standard Model value at 68\% and 95\% confidence levels with 
and without systematic error. 
On the ground of comparison we give the limits  from 
ep collider in Table \ref{tab5} using 
Weizs\"{a}cker-Williams approximation.

From these tables we see that
$\gamma p$ mode of LC+HERAp probes $\Delta\kappa$ and 
$\lambda$ with better sensitivity than present colliders and 
comparable with LHC in the case of $\Delta\kappa$ but worse than
linear $e^{-}e^{+}$ collider. The advantage of the process 
$\gamma p\to Wj$ is that it probes the $WW\gamma$ couplings 
independently of $WWZ$ effects. After further improvement of
energy and luminosity, linac-ring type $\gamma p$ collider
possibly will give complementary information to LHC and LC.

\vskip 2.0cm

\begin{table}[bth]
\caption{Sensitivity of LC+HERAp based $\gamma p$ collider to
anomalous couplings for real photons. $L_{int}=200pb^{-1}$.
Only one of the couplings is assumed to deviate from the SM
at a time.\label{tab4}}
\begin{center}
\begin{tabular}{lcc}
Sys. error & 68\%C.L. & 68\%C.L. \\
\hline
$\delta^{sys}=0$ & -0.019$<\Delta\kappa<$0.019 &
-0.075$<\lambda<$0.075 \\
$\delta^{sys}=0.02$ & -0.022$<\Delta\kappa<$0.022 &
-0.078$<\lambda<$0.078 \\
\hline
& 95\%C.L. & 95\%C.L. \\
\hline
$\delta^{sys}$=0 & -0.038$<\Delta\kappa<$0.037 &
-0.11$<\lambda<$0.11 \\
$\delta^{sys}=0.02$ & -0.044$<\Delta\kappa<$0.042 &
-0.11$<\lambda<$0.11 \\
\end{tabular}
\end{center}
\end{table}

\begin{table}[bth]
\caption{Sensitivity of LC+HERAp  collider to
anomalous couplings for WWA. $L_{int}=200pb^{-1}$.
\label{tab5}}
\begin{center}
\begin{tabular}{lcc}
Sys. error & 68\%C.L. & 68\%C.L. \\
\hline
$\delta^{sys}=0$ & -0.068$<\Delta\kappa<$0.066 &
-0.28$<\lambda<$0.28 \\
$\delta^{sys}=0.02$ & -0.069$<\Delta\kappa<$0.067 &
-0.28$<\lambda<$0.28 \\
\hline
& 95\%C.L. & 95\%C.L. \\
\hline
$\delta^{sys}$=0 & -0.14$<\Delta\kappa<$0.13 &
-0.39$<\lambda<$0.39 \\
$\delta^{sys}=0.02$ & -0.14$<\Delta\kappa<$0.13 &
-0.39$<\lambda<$0.39 \\
\end{tabular}
\end{center}
\end{table}


\acknowledgements
Authors are grateful to the members of the AUHEP group 
and S. Sultansoy for drawing our attention to anomalous 
couplings.

\end{document}